\documentclass[aps,prl,twocolumn,superscriptaddress,showpacs,longbibliography]{revtex4-1}

\usepackage{amsmath}
\usepackage{graphicx}
\usepackage{amssymb}
\usepackage[T1]{fontenc}
\usepackage[utf8]{inputenc}
\usepackage{hyperref}
\usepackage{color}
\usepackage{float}

\newcommand{\vect}{\boldsymbol}
\newcommand{\mat}{\mathrm}
\newcommand{\ie}{\mathrm i}

\definecolor{meinmagenta}{rgb}{0.8, 0.16, 0.56}
\definecolor{meinorange}{rgb}{1, 0.27, 0}
\definecolor{meingruen}{rgb}{0, 0.55, 0.27}
\definecolor{meinhellblau}{rgb}{0, 0.75 ,1}
\definecolor{meindunkelblau}{rgb}{0, 0, 0.55}

\begin{document}


\title{Self-Organized Synchronization and Voltage Stability in Networks of Synchronous Machines}


\author{Katrin Schmietendorf}
\email{katrin.schmietendorf@uni-oldenburg.de}
\affiliation{Carl-von-Ossietzky-Universit\"at Oldenburg, Institut f\"ur Physik, TWiST, 26111 Oldenburg}
\affiliation{Westf\"alische Wilhelms-Universit\"at M\"unster, Institut f\"ur Theoretische Physik, 48149 M\"unster}
\author{Joachim Peinke}
\affiliation{Carl-von-Ossietzky-Universit\"at Oldenburg, Institut f\"ur Physik, TWiST, 26111 Oldenburg}
\author{Rudolf Friedrich}
\affiliation{Westf\"alische Wilhelms-Universit\"at M\"unster, Institut f\"ur Theoretische Physik, 48149 M\"unster}
\author{Oliver Kamps}
\affiliation{Center for Nonlinear Science, 48149 M\"unster}

\date{\today}

\begin{abstract}
The integration of renewable energy sources in the course of the energy transition is accompanied by grid decentralization and fluctuating power 
feed-in characteristics. This raises new challenges for power system stability and design. 
We intend to investigate power system stability from the viewpoint of self-organized synchronization aspects.
In this approach, the power grid is represented by a network of synchronous machines.
We supplement the classical Kuramoto-like network model, which assumes constant voltages, with dynamical voltage equations, and
thus  obtain an extended version, that incorporates the coupled  categories voltage stability and rotor angle synchronization.
We compare disturbance scenarios in small systems simulated on the basis of both classical and extended model
and we discuss resultant implications and possible applications to complex modern power grids.
\end{abstract}


\maketitle


\section{Introduction}

The progressive grid integration of renewable energy plants implies substantial changes
concerning both grid topology and feed-in characteristics.
A centralized grid with unidirectional power flow from a few large conventional production units to the consumers
via levels of decreasing voltage is being replaced by a decentralized or distributed
grid, i.\,e. mainly small and medium power plants connected to the medium and low voltage levels are geographically localized near 
the consumers. Besides, high-output generating units or assemblies like offshore wind parks require long-range
transmission lines, which are able to transport large amounts of power to distant consumers.
Furthermore,  the preferred renewable energy technologies wind and solar  display fluctuations on various time scales and therefore
pose a novel challenge for grid stability. Their power input is predictable only to a limited extent and can not be customized to the current 
demand. In a future \textit{smart grid} intelligent producers, consumers and storages communicating with each other
and adapting to the grid's actual situation will form a highly complex power system.
In view of this development, power grid stability and design are actual key issues.
\\
\\
We are going to address the question of power system stability to networks of coupled synchronous machines, which are prototypes of systems 
converting mechanical power into electrical power (generators) and vice versa (motors).
A power grid is a complex dynamical system, that is constantly subjected to small disturbances such as 
small changes in production or demand and, at times, to severe disturbances, e.\,g.  failures of generating units, loads or transmission lines.
Power system stability is defined as the grid's ability to regain the former or another acceptable operating equilibrium
after a particular disturbance. 
It can be categorized into \textit{voltage stability} and \textit{rotor angle stability}.
The former is associated with constant voltages at all nodes.
The latter means the ability of synchronous machines to
remain in synchronism after a disturbance \cite{kundur,ieeetaskforce}. \\
\\
There is a notable relationship between
power system stability and synchronization phenomena in nonlinear dynamics of coupled interacting
subsystems as the synchro-nous machine's dynamical equations 
can be shown to correspond to a modified version of the prominent Kuramoto model (KM) \cite{filatrella}.
Synchronization processes occur in various fields: from heart cells or neurons to swarms of fireflies or clapping audiences to Josephson
junctions \cite{acebron05,kurths}.\\
The KM describes the behaviour of a population of coupled phase oscillators.
Its original form with equally weight-ed all-to-all coupling reads
\begin{equation}\label{eq:KM}
 \dot{\theta}_i=\omega_i+\frac KN\sum_{j=1}^N \sin(\theta_j-\theta_i)\quad\quad (i=1,...,N)
\end{equation}
with $\theta_i$ being the $i$-th oscillator's phase, $\omega_i$ its natural frequency drawn from a unimodal distribution $g$($\omega$)
and $K$ the coupling strength regulating the oscillators' interaction. In the mean-field case ($N\rightarrow\,\infty$), at a critical coupling
value $K_c=\frac{2}{\pi g(0)}$, the model displays
a phase transition from incoherence to partially synchronized states meaning that 
a group of oscillators whose natural frequencies are located
near the centre of $g$($\omega$) runs at the same frequency with constant
phase shifts \cite{kurths,strogatz00,acebron05}. Several modifications of eq. (\ref{eq:KM}) have been investigated,
e.\,g. additional inertia terms, multimodal and non-symmetric distributions
$g(\omega)$, different types of noise, various coupling szenarios, external fields and time-delayed coupling 
(\cite{kurths,acebron_KM_inertia,acebron_adaptive,acebron_uncertainty,pazo08,acebron98}, for an overview and reference to further literature see 
\cite{acebron05}).\\
In consideration of the subject's topicality, the connection between power system stability and synchronization
phenomena described by the KM has aroused only slight attention in engineering as well as nonlinear dynamics
communities (except e.\,g. \cite{dorflerSyncTransStab10,filatrella,susuki,timme12}).\\
\\
The above-mentioned categorization into rotor angle and voltage stability is
rather formal. De facto, both types of stability are coupled and instabilities often emerge mutually \cite{ieeetaskforce}. 
However, the Kuramoto-like machine representation
of the \textit{classical model}, which has been the means of choice for the investigation of power system stability from nonlinear dynamics 
research so far, assumes constant voltages. In order to involve both stability categories, we start out from a more
detailed synchronous machine model, which takes into account its electrodynamical behaviour. This yields a more 
realistic, but still highly reduced network model.
In contrast to the \textit{classical model}, the resulting \textit{extended model} includes dynamical equations for the nodal voltages and the important
feature of voltage-angle stability interplay. Comparing the system behaviour modeled by the classical and the extended
equations indicates significantly different stability predictions for certain disturbance scenarios.\\
\\
The paper is organized as follows:\\
In section I we outline the derivations of the \textit{classical model} and the \textit{extended model}. 
First we introduce into the graph theoretical representation of an electrical network by means of the \textit{nodal 
admittance matrix} plus the \textit{power flow equations} (see subsection I.a) and present the \textit{swing equation} governing the synchronous machine's mechanical dynamics (see subsection I.b). 
As an interim conclusion, in subsection I.c we arrive at dynamical equations associated with the classical model, which uncover the 
relationship to synchronization phenomena described by the KM. We briefly discuss the classical model's main shortcomings.
Subsequently, in subsection I.d we sketch the extension of the classical model by dynamical equations for the machines' voltages.
This yields a novel type of KM modification,  which has not been investigated in the context of theoretical nonlinear sciences yet.
In section II we present numerical simulations of small systems during and after certain disturbances both in the classical and in the extended 
representation. As a last point, the consequences and potential applications for stability investigations on networks of synchronous machines
 modeling modern power grids with high percentage of renewables are discussed in section III.

\section{I The Model}\label{sec:model}

\subsection{I.a Network Representation and Power Flow}\label{sec:representation_powerflow}
An electrical power grid's elements form a complex graph $G$($V,E$), i.\,e. a network. The set of nodes $V$ consists of
production units, loads, transformers, intersection points etc., the edges or links $E$ correspond to transmission lines.
Consider an electrical network consisting of $|V|$=$M$ nodes. Each link ($k$,\,$l$) ($k$,\,$l$\,$\in\{1,...,M\}$) is weighted by
a complex-valued admittance $Y_{kl}=G_{kl}+\ie B_{kl}$ ($G_{kl}$: conductance, $B_{kl}$: susceptance). Kirchhoff's and Ohm's laws yield the 
\textit{nodal network equations} \cite{machowski,kundur}:
\begin{equation}
 \vect I= \mat Y_\text{net}\vect V\,.
\end{equation}
$\vect I$ and $\vect V$ are the vectors of the complex nodal voltages $V_j$ and currents $I_j$ ($j\,=\,1,...,M$) and 
$\mat Y_\text{net}\in\mathbb C^{M\times M}$ is the 
\textit{(nodal) admittance matrix}. $\mat Y_\text{net}$ corresponds to the network's Laplacian matrix
\begin{equation}
 \mat Y_\text{net}=\mat L_\text{net}=\mat G_\text{net}-\mat A_\text{net}\,,
\end{equation}
$\mat A_\text{net}\in\mathbb C^{M\times M}$ being the \textit{weighted adjacency matrix} with coefficients
\begin{equation}
 a_{kl}=\begin{cases} Y_{kl} & \text{if nodes $k$ and $l$ are connected by $Y_{kl}$}\\ 0 & \text{else}\end{cases}
\end{equation}
and $\mat G_\text{net}\in\mathbb C^{M\times M}$ being the \textit{diagonal degree matrix} whose element $d_{kk}$ equals the sum of 
admittances linked to node $k$ \cite{tittmann}.
\textit{Passive} nodes with $I_j$=\,0 (intersections, loads modeled by passive admittances and suchlike) can be eliminated via \textit{Kron reduction} of 
$\mat Y_\text{net}$ and accordingly by reduction of the corresponding network \cite{machowski,kundur}.
This leads to a well-definded \textit{reduced admittance matrix} $\mat Y_\text{red}=:\mat Y
\in\mathbb C^{N\times N}$ 
being a reduced network's Laplacian in turn and relating the nodal currents and voltages of the $N<M$ \textit{active} ($I_j\neq$\,0) nodes
(for a detailed discussion of the Kron reduction of matrices and their implications for the corresponding graphs 
see \cite{dorflerkron11}).\\
The apparent power at node $j$ reads
\begin{equation}
 S_j=V_j I^*_j
\end{equation}
with $V_j=|V_j|e^{\ie\delta_j}$ and $I_j=\sum_k Y_{jk}V_k$ ($\delta_j$: electrical phase angle). 
Substituting $Y_{jk}=|Y_{jk}|e^{\ie\theta_{jk}}=G_{jk}+\ie B_{jk}$ (with $Y_{jk}$ now being the $jk$-component of the
admittance matrix) yields
the real power $P_j=\operatorname{Re}(S_j)$ 
\begin{equation}\label{eq:powerflow}
 P_j=\sum_{k=1}^N |V_j||V_k|\big[G_{jk}\cos(\delta_j-\delta_k)+B_{jk}\sin(\delta_j-\delta_k)\big]
\end{equation}
\cite{machowski,kundur}.
Note the power flow's dependence on the phase angle differences between node $j$ and its adjacent nodes.

\subsection{I.b The Swing Equation}\label{sec:swing}

\begin{figure}[H]
 \includegraphics[scale=0.6]{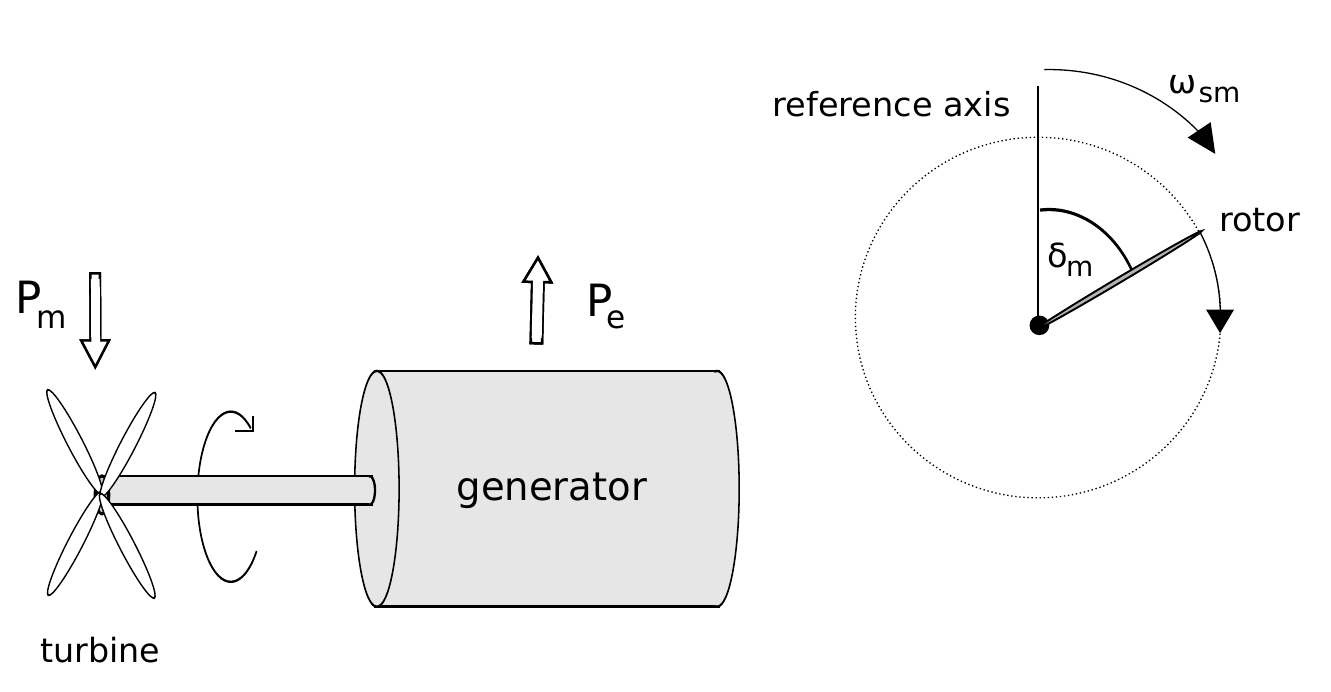}
\caption{Schematic picture of a synchronous generator and its mechanical phase angle $\delta_\text m$ with respect to a rotating reference axis.}
\label{fig:swing}
\end{figure}

Synchronous generators convert the mechanical input pow-er of their turbine $P_\text{m}>$\,0 into electrical power $P_\text{e}$
(see fig.\,\ref{fig:swing}).
They owe their name to the synchronicity of the rotating magnetic field of the rotor and the alternating voltages and currents
induced in the stator windings (for construction and functionality of synchronous machines see \cite{kundur,machowski}).
The mechanical rotor angle $\delta_\text m$ denotes the angular difference between the rotor axis
and a reference axis rotating with synchronous angular velocity $\omega_\text{sm}$ (system frequency) (see fig.\,\ref{fig:swing}).
The \textit{swing equation}\cite{kundur,machowski} governing the rotor's mechanical dynamics is derived from Newton's law for rotating masses.
It reads
\begin{equation}\label{eq:swing}
 M_\text m\frac{\mathrm d^2\delta_\text m}{\mathrm dt^2}+D_\text m\frac{\mathrm d\delta_\text m}{\mathrm dt}=P_\text m-P_\text e
\end{equation}
($M_\text m$: angular momentum at $\omega_\text{sm}$; $D_\text m$: damping torque at $\omega_\text{sm}$). The electrical phase angle $\delta$ and mechanical angle are related by $\delta=\frac{2\delta_\text m}{p}$ with $p$ being the generator's number of magnetic poles. In case of
a two-pole generator both angles are identical.\\
These explanations concerning synchronous generators can be transferred to synchronous motors with electrical and mechanical powers changing 
parts, i.\,e. electrical power is converted into mechanical power ($P_\text{m}<$\,0).

\subsection{I.c The Classical Model}\label{sec:classical_model}
Combining eq. (\ref{eq:powerflow}) and (\ref{eq:swing}) on the assumption of a lossless network ($G_{ij}$=\,0 $\forall\,i,j$) yields for the the $i$-th machine's dynamics 
\begin{equation}\label{eq:classical_model}
 M_i\ddot \delta_i= -D_i\dot\delta_i +P_{\text{m},i}-\sum_{j=1}^N V_iV_jB_{ij}\sin(\delta_i-\delta_j)
\end{equation}
with $V_i$ being its voltage amplitude and $P_{ij}=V_iV_jB_{ij}$ the maximum transferred power between 
machines $i$ and $j$ (for an alternative derivation of eq. (\ref{eq:classical_model}) based on a power balance equation see
\cite{filatrella}).
Note that eq. (\ref{eq:classical_model}) corresponds to a modification of the KM eq. (\ref{eq:KM}) with additional inertia terms.\\
Eq. (\ref{eq:classical_model}) is associated with the \textit{classical model} 
(see \cite{kundur,machowski,dorflerSyncTransStab10}), which
implies certain further assumptions being shortcomings in some respects \cite{anderson}. 
For instance, one assumes constant voltages $V_i$ and constant mechanical power $P_{\text m,i}$. 
The former makes the model incapable of modeling voltage dynamics or angle-voltage stability interplay.
The latter conflicts with fluctuating feed-in $P_{\text m,i} (t)$, especially considering the characteristics of wind and solar power plants.
Furthermore, in the classical representation all loads are modeled by constant impedances, which, as they
are passive nodes, can be eliminated. Here we choose synchronous motors ($P_{\text m,i}<0$) as loads instead. This allows for the fact that 
loads have their individual temporal dynamics. Synchronous motors are modeled analogous to synchronous generators (c.\,f. subsection I.b).

\subsection{I.d The Extended Model Including Voltage Dynamics}\label{sec:improved_model}
The extended model including voltage dynamics is based on a more detailed synchronous machine representation,
that takes into account the machine's electrodynamical behaviour to a certain extent. Its derivation starts at the basic equations governing
the electromagnetical interactions between the involved field, damping and stator windings given in the $abc$-stator reference system ($a$,$b$ and $c$ denoting the three stator phases).
The final model is formulated in $dq$-rotor coordinates with the $d$-axis centered in the rotor field's magnetic north pole and the $q$-axis 
perpendicular to it (for a detailed discussion of the following derivation including $abc\rightarrow dq(0)$ transformation see \cite{machowski}, also \cite{kundur}).
One distinguishes into three characteristic machine states: subtransient, transient and stationary.
The generator after a disturbance is modeled by subtransient and transient voltages behind respective reactances. The governing equations for the subtransient voltages
$E_{d/q}''$ and the transient voltages $E_{d/q}'$ read
\begin{eqnarray}\label{eq:6order}
 T_{d0}''\dot E_q''&=&E_q'-E_q''+I_d(X_d'-X_d'')\, ,\notag \\
 T_{q0}''\dot E_d''&=&E_d'-E_d''-I_q(X_q'-X_q'')\, ,\notag \\
 T_{d0}'\dot E_q'&=&E_\text f-E_q'+I_d(X_d-X_d')\, ,\notag\\
T_{q0}'\dot E_d'&=&-E_d'-I_q(X_q-X_q')
\end{eqnarray}
($T_{d0/q0}'$, $T_{d0/q0}''$: transient/subtransient time constants of the $d$- and $q$-axis; 
$X_{d/q}'$, $X_{d/q}''$: transient/subtransient reactances;
 $I_d$, $I_q$: armature currents; $E_\text f\sim$\,rotor's field voltage). 
The machine's representation is completed by the swing equation (\ref{eq:swing}).
Neglecting damper winding effects and setting $E_d'$=\,0 and $X_q'$=$X_q$ reduces eq. (\ref{eq:swing}) and  (\ref{eq:6order})
to the \textit{third-order-model} for the $i$-th machine:
\begin{eqnarray}\label{eq:modell3}
    M_i\ddot\delta_i&=&-D_i\dot\delta_i+P_{\text m,i}-P_{\text e,i}\, ,\notag\\
  T_{d0,i}'\dot E_{q,i}'&=&E_{\text f,i}-E_{q,i}'+I_{d,i}(X_{d,i}-X_{d,i}')\, .
\end{eqnarray}
The electrical power (assuming $X_d'=X_q'$) is
\begin{equation}
 P_{\text e,i}=3(E_{d,i}'I_{d,i}+E_{q,i}'I_{q,i})\,.
\end{equation}
Using the relationship between the $i$-th machine's individual ($dq$) rotor coordinates and complex ($ab$) network coordinates 
\begin{equation}
 \left(\begin{array}{c}
I_{d,i}\\I_{q,i}
\end{array}\right)
= \left(\begin{array}{cc}
-\sin\delta_i & \cos\delta_i \\ \cos\delta_i & \sin\delta_i
\end{array}\right)
\cdot\left(\begin{array}{c}
I_{a,i}\\I_{b,i}
\end{array}\right)\,
\end{equation}
($E'_{d,i},E'_{q,i}\rightleftarrows E'_{a,i},E'_{b,i}$ analogous) one can write the $d$-axis current and electrical power as follows
(with $\delta_{ij}:=\delta_i-\delta_j$):
\begin{eqnarray}\label{eq:I_di}
 I_{d,i}=\sum_{j=1}^N \left(G_{ij}\cos\delta_{ij}+B_{ij}\sin\delta_{ij}\right)E_{d,j}^\prime\notag\\
-\left(G_{ij}\sin\delta_{ij}-B_{ij}\cos\delta_{ij}\right)E_{q,j}^\prime\,,
\end{eqnarray}
\begin{eqnarray}\label{eq:P_ei}
 P_{\text e,i}&=&3(E_{d,i}^\prime I_{d,i}+E_{q,i}^\prime I_{q,i})\notag\\
 &=&3E_{d,i}^\prime\Big[\sum_{j=1}^N \left(G_{ij}\cos\delta_{ij}+B_{ij}\sin\delta_{ij}\right)E_{d,j}^\prime\notag\\
&+&\left(-G_{ij}\sin\delta_{ij}+B_{ij}\cos\delta_{ij}\right)E_{q,j}^\prime\Big]\notag\\
&+&3E_{q,i}^\prime\Big[\sum_{j=1}^N \left(G_{ij}\sin\delta_{ij}-B_{ij}\cos\delta_{ij}\right)E_{d,j}^\prime\notag\\
&+&\left(G_{ij}\cos\delta_{ij}+B_{ij}\sin\delta_{ij}\right)E_{q,j}^\prime\Big]\,.
\end{eqnarray}
Assuming a lossless network plus factoring in $E_d'$=\,0 simplifies eq. (\ref{eq:I_di}) and (\ref{eq:P_ei})
and finally yields
\begin{eqnarray}\label{eq:modell_final}
 M_i\ddot\delta_i&=& -D_i\dot\delta_i+P_{\text m,i}(t)-\sum_{j=1}^N B_{ij}E_{q,i}^\prime E_{q,j}^\prime\sin\delta_{ij}\, ,\notag\\
T_{d0,i}\dot E_{q,i}^\prime&=&E_{\text f,i}-E_{q,i}^\prime+(X_{d,i}-X_{d,i}^\prime)\sum_{j=1}^N B_{ij}E_{q,j}^\prime\cos\delta_{ij}\,.\notag\\
\end{eqnarray}
Eq. (\ref{eq:modell_final}) enormously reduces the dynamics of a network of synchronous machines, but 
the model still includes the features of voltage dynamics and angle-voltage stability interplay. The susceptance matrix coefficients
$B_{ij}$ allow for variations concerning the network's topology. The mechanical input/output power $P_{\text m,i}(t)$ can be fitted to the
production units' (generators) and consumers' (motors) characteristics.\\
The extended model eq. (\ref{eq:modell_final}) can be interpreted as a modification of the KM
\begin{eqnarray}\label{{eq:KM_vgl}}
m_i\ddot\theta_i+D_i\dot\theta_i=\omega_i+\sum_{j=1}^N K_{ij}(\{E'_{q,k}(t)\})\sin\delta_{ij} +\xi_i(t)\notag\\
\end{eqnarray}
with additional inertia and
optional noise terms $\xi_i(t)$ depending on the specific choice of $P_{\text m,i}(t)$. Inertia and noise in the KM have already been considered, at least
for several specific cases (for references see Introduction). However, this particular type of time-dependent coupling coefficients 
$K_{ij}(t)=B_{ij}E_{q,i}^\prime(t) E_{q,j}(t)$, whose dynamics depend on the oscillators' phase differences in turn, is a novel type of KM modification and has not been 
investigated within the scope of synchronization of coupled oscillators to the best of our knowledge. Eq.\,(\ref{eq:modell_final}) has a higher dimension than the classical model 
eq.\,(\ref{eq:classical_model}), which can lead to a different and richer system behaviour.

\section{II Simulations}
\label{sec:simulations}
\begin{figure}[h!]
\begin{center}
\includegraphics[scale=0.6]{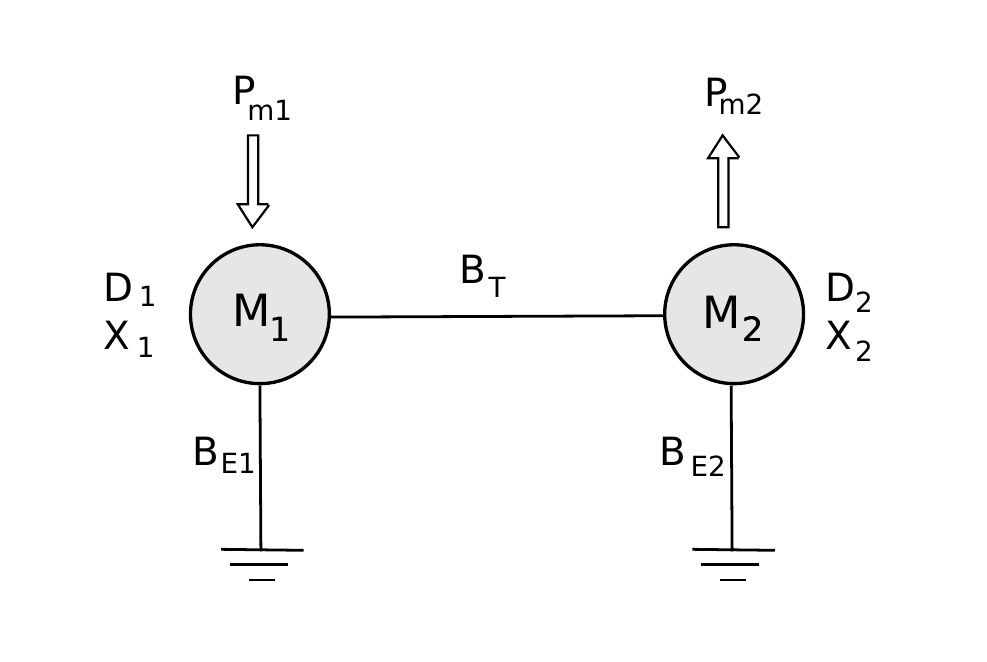}
\caption{Two-machine system of a generator M$_1$ ($P_{\text m1}>0$) and a motor M$_2$ ($P_{\text m2}<0$) linked by a line susceptance $B_T$ 
and grounded by shunt susceptances $B_{E1}$ and $B_{E2}$.}
\label{fig:twomachine}
\end{center}
\end{figure}

\begin{figure*}
 \includegraphics[scale=0.85]{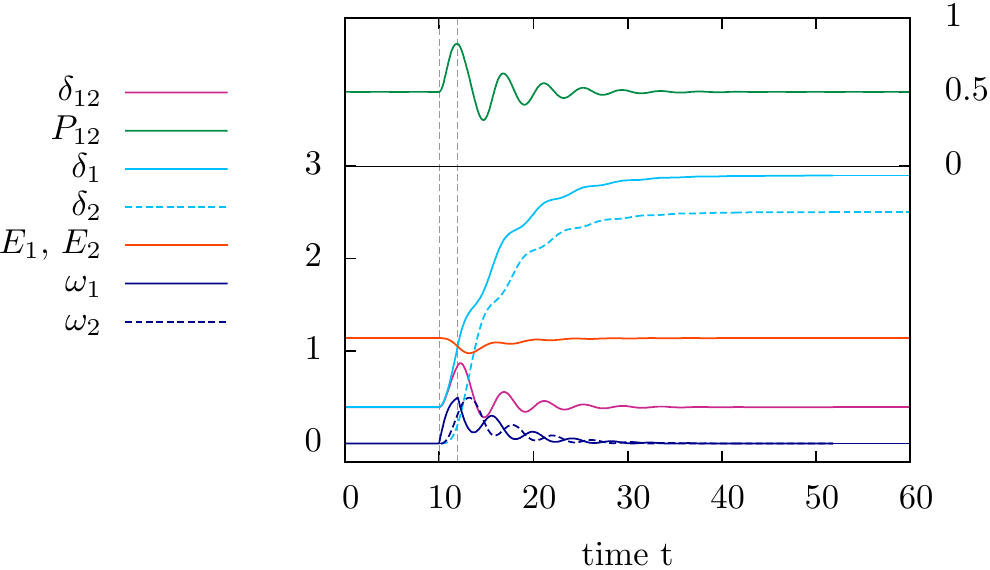}
\hspace*{2cm}
\includegraphics[scale=0.85]{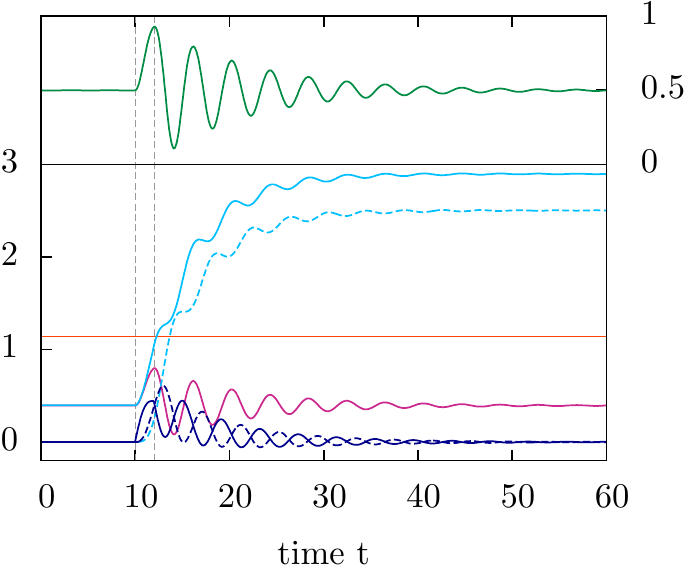}
\caption{Two machine-system. Phase angles \textcolor{meinhellblau}{$\delta_1$} (solid), \textcolor{meinhellblau}{$\delta_2$} (dashed), 
phase difference \textcolor{meinmagenta}{$\delta_{12}$}, 
angular velocities \textcolor{meindunkelblau}{$\omega_1$} (solid), \textcolor{meindunkelblau}{$\omega_2$} (dashed),
voltages \textcolor{meinorange}{$E_1$}, \textcolor{meinorange}{$E_2$}
and power transfer \textcolor{meingruen}{$P_{12}$} from $M_1$ to
$M_2$ as functions of time. $\gamma_1=\gamma_2=0.2$, $P_{\text m1}=-P_{\text m2}=0.5$, $\alpha_1=\alpha_2=2.0$, $E_{\text f,1}=E_{\text f,2}=1.0$,
$X_1=X_2=1.0$, $B_{11}=B_{22}=-0.8$, $B_{12}=B_{21}=1.0$. $P_\text{dist}=1.0$ during $t\in[10,12]$ (disturbance period denoted by vertical dotted lines).
Left: Extended model. Right: Classical model. The machine voltages \textcolor{meinorange}{$E_1$} and \textcolor{meinorange}{$E_2$} are congruent because of identical machine and line parameters. Both systems return to stationary operation.}
\label{fig:two-machine_1}
\end{figure*}

\begin{figure*}
 \includegraphics[scale=0.85]{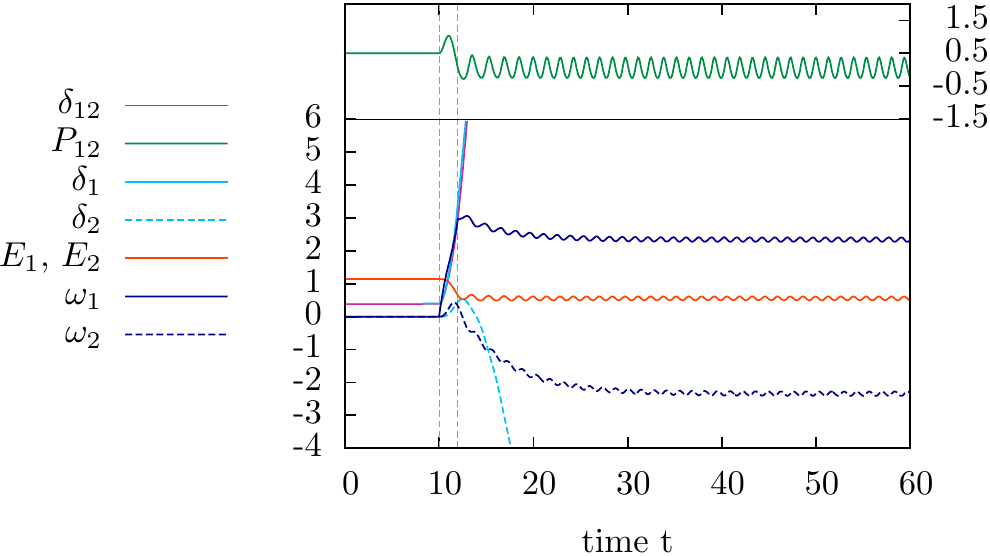}
\hspace*{2cm}
\includegraphics[scale=0.85]{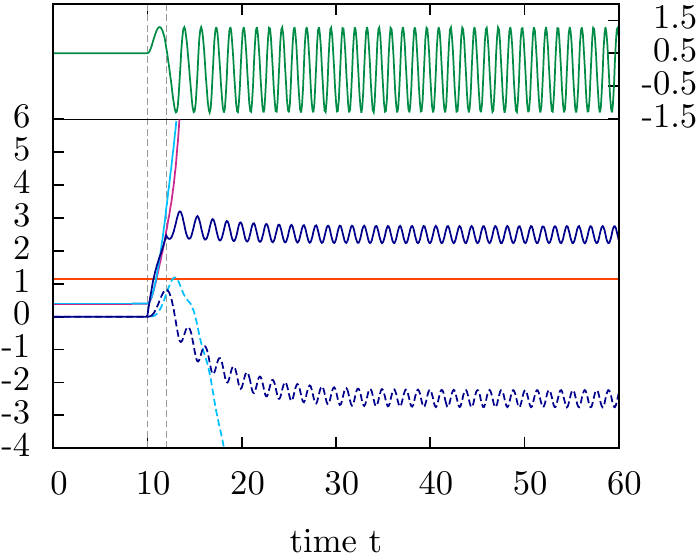}
\caption{Two-machine system. Disturbance scenario with $P_\text{dist}=2.5$\,. The other parameter values as in fig.\,\ref{fig:two-machine_1}.
Left: Extended model. \textcolor{meinhellblau}{$\delta_1$}, \textcolor{meinmagenta}{$\delta_{12}$} (\textcolor{meinhellblau}{$\delta_2$}) display unbounded growth (decrease) beyond the shown interval. Right: Classical model. Both systems are unstable in the sense of power system stability.}
\label{fig:two-machine_3}
\end{figure*}

Consider the normalized $N$-machine system (after renaming the normalized quantities):
\begin{eqnarray}\label{eq:model_final_norm}
 \ddot\delta_i&=&-\gamma_i\dot\delta_i+P_{\text m,i}-\sum\limits_{j=1}^N B_{ij}E_iE_j\sin\delta_{ij}\,,\notag\\
 \alpha_i\dot E_i&=&E_{\text f,i}-E_i+X_i\sum\limits_{j=1}^N B_{ij}E_j\cos\delta_{ij} 
\end{eqnarray}
with $B_{ij}<0$ for $i=j$, $B_{ij}>0$ for $i\neq j$ and $X_i>0$ being generally valid.
We consider the cases of a two-machine system as the basic component of complex power grids and a six-machine system referring to the Zealand power grid approximation in \cite{filatrella}. Both systems are subjected to temporary disturbances. We compare the system behaviour based on model eq. (\ref{eq:model_final_norm}) with the behaviour predicted by the more reduced
classical description, which is obtained by restricting to the first line of eq. (\ref{eq:model_final_norm}) and assuming constant
voltages.\\
\\
\textbf{Two-Machine System}\\
\\
First we consider the case of a $N$=2\,-machine system consisting of a generator connected to a motor (see fig.\,\ref{fig:twomachine})
with symmetric lines and identical machine parameters.\\
\\ 
The situations depicted in fig.\,\ref{fig:two-machine_1} to fig.\,\ref{fig:two-machine_3} are as follows: The systems
(parameters denoted in fig.\,\ref{fig:two-machine_1}) are in steady
states with constant phase angles $\delta_1^*$, $\delta_2^*$
and phase angle difference $\delta_{12}^*=0.395$, $\omega_1^*=\omega_2^*=0$ (meaning that the machines run with system frequency, cf. I.b),
constant voltages $E_1^*=E_2^*=1.140$ (in the classical system the voltages are parameters, whose values are determined by the stationary
values of the extended model)
and stationary power transfer $P_{12}^*=B_{12}E_1^*E_2^*\sin\delta_{12}^*=0.5=P_{\text m,1}$ from the generator to the 
motor. In other words, the system is in its fixed point.
The injected and the consumed power match: $\sum_{i=1,2}P_{\text m,i}=0$.
During the denoted time interval $t\in[10,12]$ the systems are subjected to certain disturbances in terms of an increase of parameter 
$P_{\text m,1}$ to $P_{\text{dist}}$, which corresponds to a temporal power feed-in plus.\\
\begin{figure*}
 \includegraphics[scale=0.85]{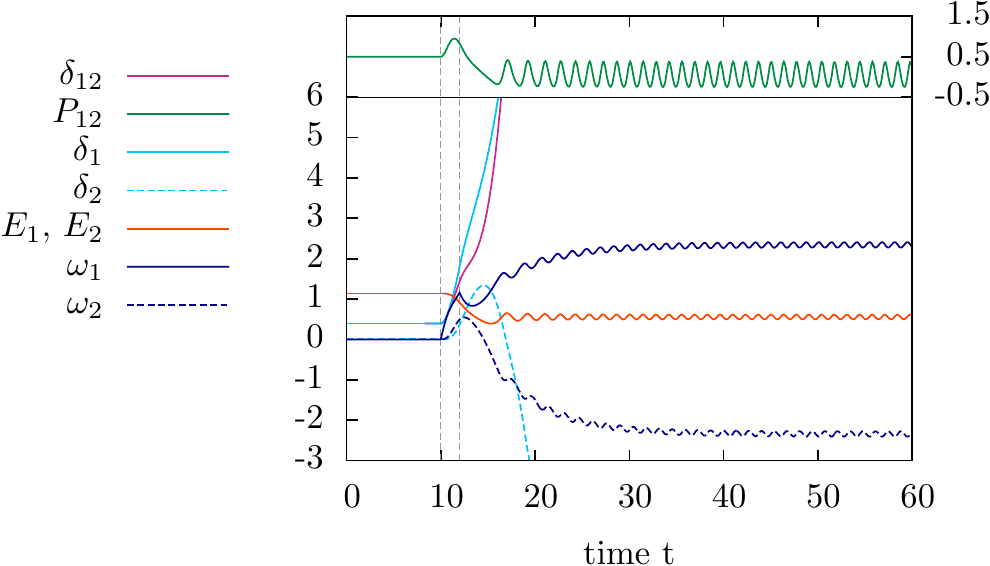}
\hspace*{2cm}
\includegraphics[scale=0.85]{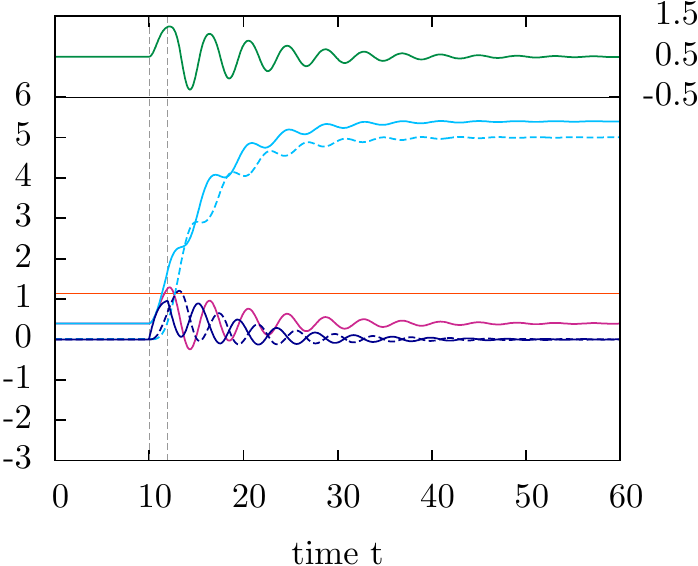}
\caption{Two-machine system. Disturbance scenario with $P_\text{dist}=1.5$\,. The other parameter values as in fig.\,\ref{fig:two-machine_1}.
Left: Extended model. \textcolor{meinhellblau}{$\delta_1$}, \textcolor{meinmagenta}{$\delta_{12}$} (\textcolor{meinhellblau}{$\delta_2$}) display unbounded growth (decrease) beyond the shown interval. Right: Classical model. Unlike the classical system, the extended system is not stable in the sense of power system stability.}
\label{fig:two-machine_2}
\end{figure*}
\\
In fig.\,\ref{fig:two-machine_1} the temporal increase of the power feed-in from $P_{\text m,1}=0.5$ to
$P_{\text{dist}}=1.0$ makes the generator accelerate and the two machines'
phase difference grow. In the extended system the nodal voltages drop, while in the classical system the voltages are constants by definition
(they are plotted yet, for the sake of consistency).
After the pertubation both systems return to steady operation, i.\,e. their initial fixed point, with decaying oscillations. 
The extended system reaches the fixed point - or which is practically more relevant: an operating status where the deviations from the 
stationary values are acceptably small - earlier than the classical system.
In fig.\,\ref{fig:two-machine_2} the systems are subjected to a larger disturbance $P_\text{dist}=1.5$\,. While the classical system returns to stationary operation
in a qualitatively similar manner to the situation before, the extended system reaches a state with unbounded growing phase angles and phase difference
and oscillating voltages and power flow. Due to the strength of the disturbance the latter gets out of the fixed point's region of attraction and 
enters a limit cycle.
The power flow between generator and motor permanently alters its direction, which
is reflected by the changing sign of $P_{12}(t)$.
The additional increase of the disturbance to $P_\text{dist}=2.5$ (see fig.\,\ref{fig:two-machine_3}) causes both systems to operate with unbounded growing phase differences
and oscillating power, angular velocities and (for the extended model) voltages, i.\,e. both systems have approached limit cycles. In the classical system the amplitudes of power and angular velocity 
oscillations are larger than in the extended system. Both systems are not stable in the sense of
power system stability presented in the introduction.\\
These example cases illustrate that, taking the same disturbance scenario as a basis, the classical and extended model can predict quantitavely and, that is the key point, qualitatively different system behaviour.
The dynamic voltage equations in eq.\,(\ref{eq:modell_final}) involve synchronous machine parameters,
which depend on the type of generator or motor. The specific choice of these parameters, of course, 
influences the system's behaviour. In the classical model these machine parameters are left out. 
Considering solely fig.\,\ref{fig:two-machine_1}, one could argue that an increase of the mechanical damping coefficient $\gamma_i$ in the classical description
can take into account electrodynamic damping effects, which are intrinsically ignored, and correct the differences to the extended model. However, looking at the scenario in fig.\,\ref{fig:two-machine_2}, this turns out to be insufficient as the qualitative differences (stable operation/fixed point - unstable operation/limit cycle) remain.\\
\\

\begin{figure}[H]
\begin{center}
 \includegraphics[scale=0.35]{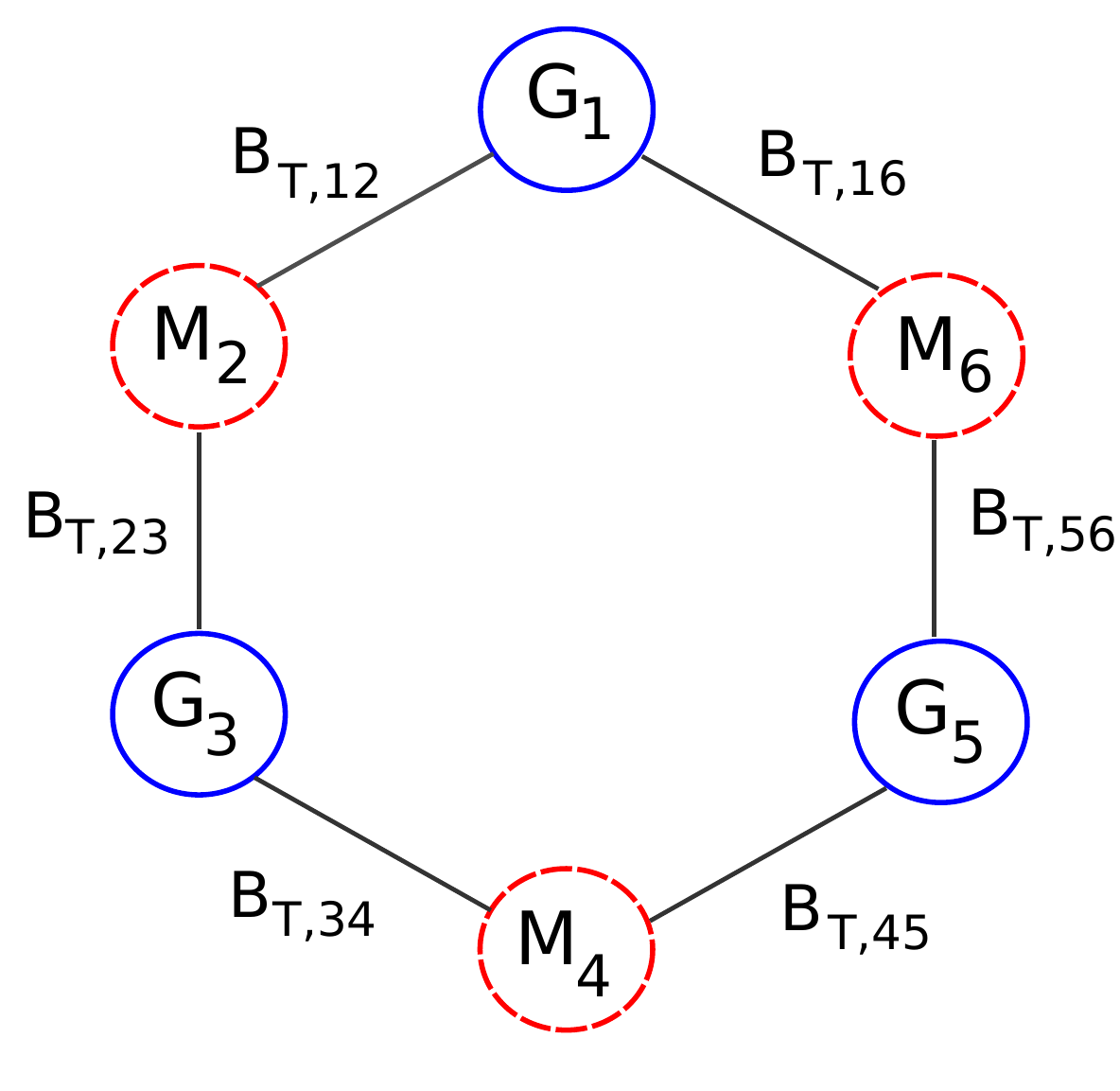}
\caption{Six-machine system of three generators \textcolor{blue}{G$_i$} ($i=$1,3,5) and three motors \textcolor{red}{M$_i$} ($i=$2,4,6). Shunt susceptances are neglected. The susceptance matrix $\{B_{ij}\}_{i,j=1,..,6}$ is calculated from the 
transfer susceptances (cf.\,\ref{sec:representation_powerflow}) $B_{\text T,12}$=\,-1.0, $B_{\text T,23}$=\,-0.5, $B_{\text T,34}$=\,-0.7, 
$B_{\text T,45}$=\,-1.0, $B_{\text T,56}$=\,-1.2, $B_{\text T,16}$=\,-0.8, $P_{\text m,1}$=0.25, $P_{\text m,2}$=\,-0.2, 
$P_{\text m,3}$=0.2, $P_{\text m,4}$=\,-1.5, $P_{\text m,5}$=1.5, $P_{\text m,6}$=\,-0.25,
$\gamma_i$=0.1 $\forall i$, the other machine parameters as in fig.\,\ref{fig:two-machine_1}.}
\label{fig:zealand}
\end{center}
\end{figure}

\begin{figure*}
 \includegraphics[scale=0.95]{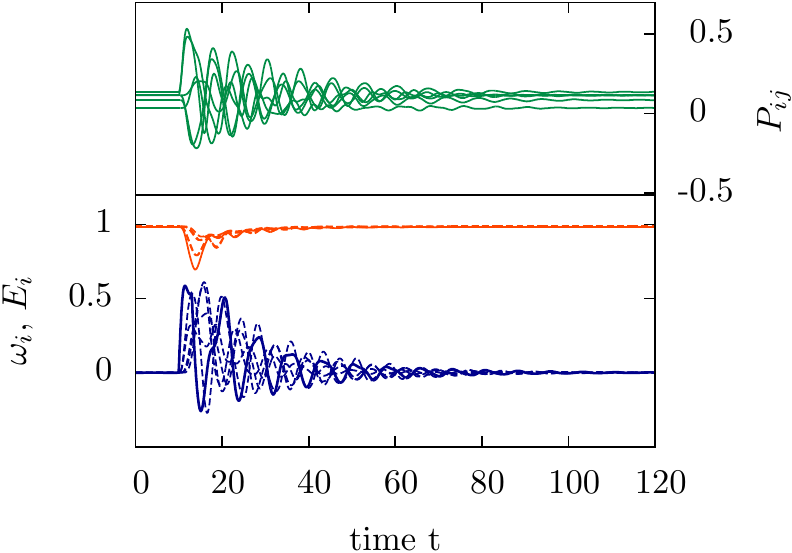}
\hspace*{1cm}
\includegraphics[scale=0.95]{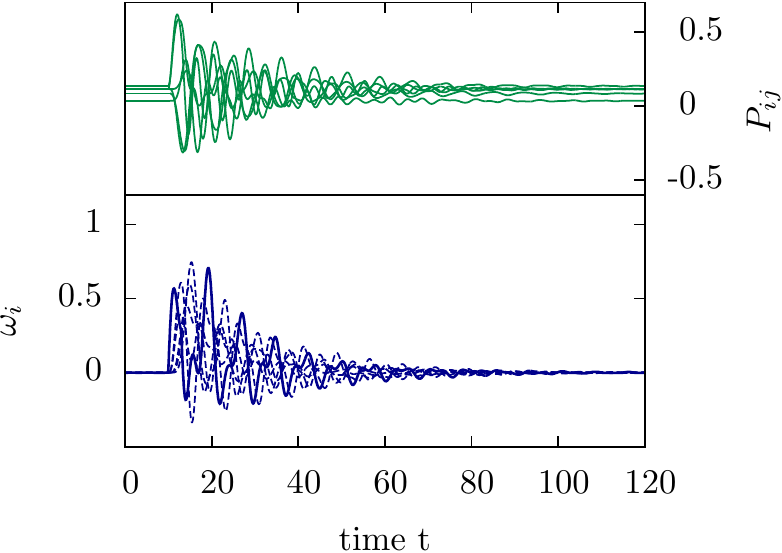}
\caption{Six-Machine system. Disturbance scenario with $P_\text{dist}=1$ for $t\in[10,13]$. Angular velocities \textcolor{meindunkelblau}{$\omega_i$}, voltages \textcolor{meinorange}{$E_i$} ($i=1$: solid line, $i=2,..,6$ 
dashed lines) and power transfer \textcolor{meingruen}{$P_{ij}$} along all links (the reference machines are chosen in such a way that all stationary
power flows are positive) as functions of time $t$. Left: Extended model. Right: Classical model. Both systems return to stationary operation.}
\label{fig:zealand_Sim1c}
\end{figure*}

\begin{figure*}
\hspace*{0.1cm}
 \includegraphics[scale=0.95]{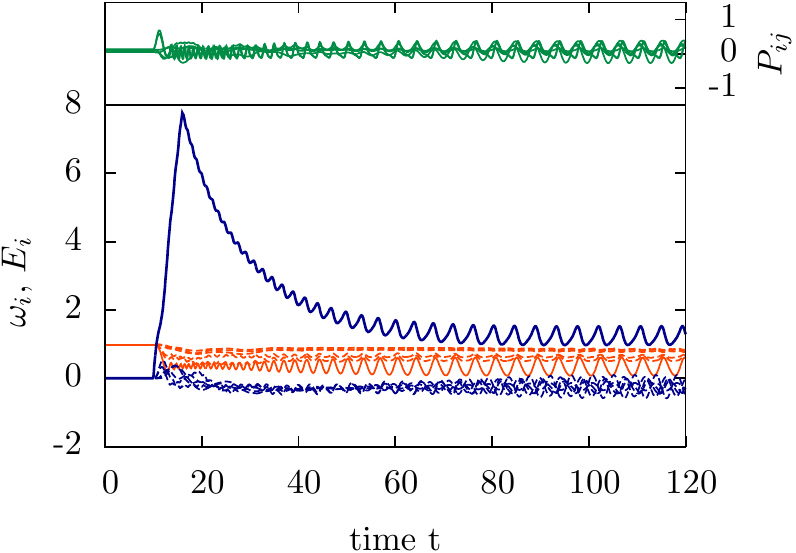}
\hspace*{1cm}
\includegraphics[scale=0.95]{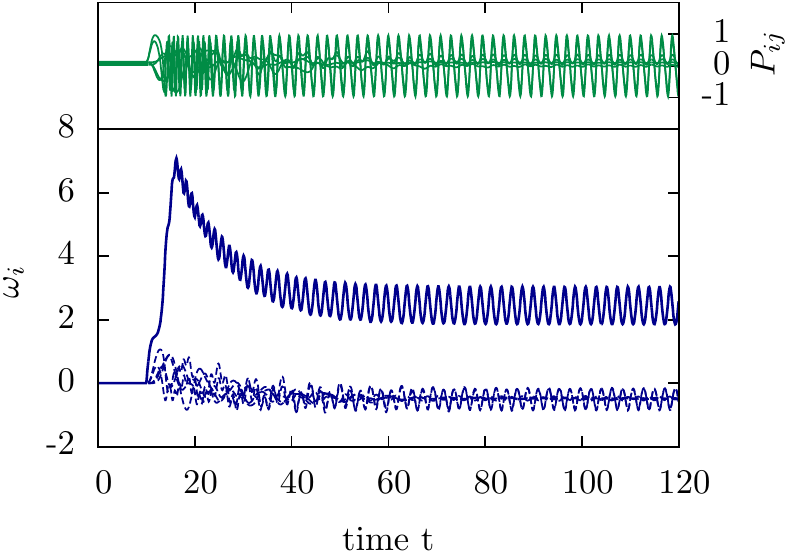}
\caption{Six-machine system. Disturbance scenario with $P_\text{dist}=3$ for $t\in[10,16]$. Left: Extended model. Right: Classical model. Both models are unstable in the sense of power system stability.}
\label{fig:zealand_Sim1b}
\end{figure*}

\noindent\textbf{Six-Machine System}\\
\\
Consider the six-machine system consisting of three generators and three machines arranged in a ring (see fig.\,\ref{fig:zealand}). 
The disturbance scenarios depicted in fig.\,\ref{fig:zealand_Sim1c}  to fig.\,\ref{fig:zealand_Sim1b} are similar to the two-machine cases. 
The stationarily operating systems are subjected to different perturbations in terms of temporary ($T_\text{dist}$) power feed-in increase at generator G$_1$ from $P_{\text m,1}$ to $P_\text{dist}$. The system behaviour of the classical framework is contrasted with the extended modeling again.\\
\begin{figure*}
\hspace*{0.1cm}
 \includegraphics[scale=0.95]{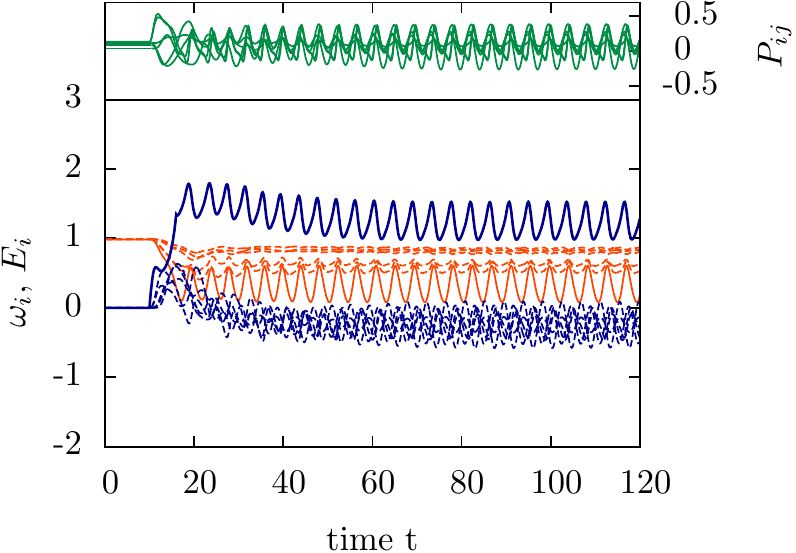}
\hspace*{1cm}
\includegraphics[scale=0.95]{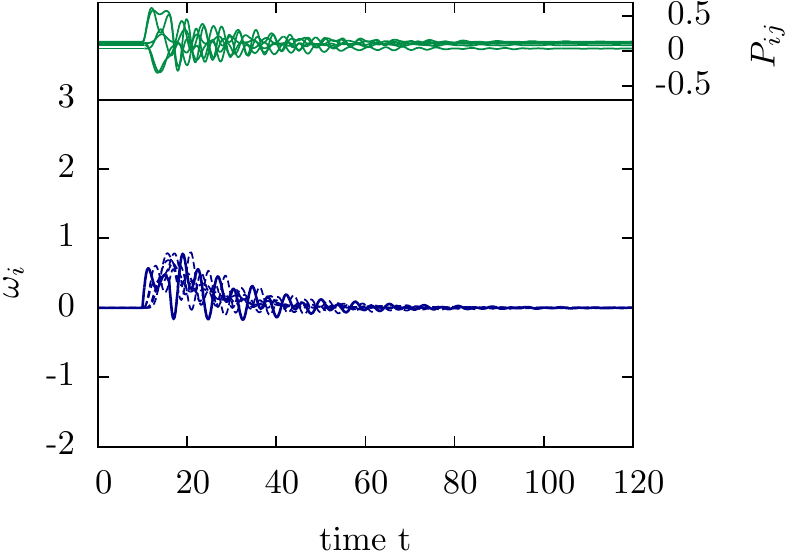}
\caption{Six-machine system. Disturbance scenario with $P_\text{dist}=1$ for $t\in[10,16]$. Left: Extended model. Right: Classical model. Unlike the classical system, the extended system is not stable in the sense of power system stability.}
\label{fig:zealand_Sim1a}
\end{figure*}
\noindent In fig.\,\ref{fig:zealand_Sim1c} ($P_\text{dist}=1$, $T_\text{dist}=3$) both systems return to stationary operation with constant
voltages, power transfer and vanishing angular velocity with respect to the system frequency.
In fig.\,\ref{fig:zealand_Sim1a} the duration of disturbance is increased to $T_\text{dist}=6$. This makes the extended system transition into
unstable operation with ocillating voltages, angular velocities and power transfers, whereas the classical system returns to stationary 
operation.
Fig.\,\ref{fig:zealand_Sim1b} shows that for an additional increase of the disturbance to $P_\text{dist}=2$ both systems pass over to
oscillating behaviour, i.\,e. they are unstable in the sense of power system stability.\\
In this case, as in the simple two-machine system, for several disturbance scenarios the extended model and the classical model yield divergent 
stability predictions. The same can be assumed for for larger networks and other topologies.\\
\\
The examples discussed above give rise to the following conclusions:
\begin{itemize}
\item [*] Both the classical and the extended model display typical features of power grid operation (see \cite{kundur}): after being subjected to a disturbance the system 
either returns to stable, synchronous operation or transitions into unstable operation with machines ''falling out of step''
(i.\,e. they no longer run at system frequency), power flow and (in the extended model) voltage oscillations and drops (for the classical framework this has already been expounded in \cite{filatrella,susuki,timme12}).
\item [*] For specific disturbance scenarios, the classical and the extended model can predict different stability behaviour. With a view to
 the investigation of complex networks of synchronous machines, the extended model should be preferred because it is more realistic and accomodates
the synchronous machines' and hence the network's electromagnetic nature to a greater extent.
\end{itemize}

\section{III Summary and Outlook}
\label{sec:summary}
We introduced an extended model for networks of synchronous machines, which exceeds the classical KM-like representation in that it includes 
voltage dynamics and the feature of voltage-angle stability interplay. It is more realistic with respect to the power grid's 
electrical character. Nonetheless, it is still highly reduced due to several simplifications (see subsection I.d) and neglecting 
active control equipment.  
For small networks there are, of course, various more accurate machine models quoted by the electrical engineering literature.
However, we intended to attain an adequate network representation for the future analysis of large networks in view of self-organisation
aspects considering synchronization \textit{and} voltage stability. Against this backdrop, the introduced model provides a promising basis
for the investigation of the consequences resulting from progressive grid integration of renewable energy sources mentioned in the introduction. On the one hand, one is able to analyze the effects of short-term feed-in fluctuations induced by wind and solar plants via choosing suitable $P_{\text m,i}(t)$ for the generating units 
($P_{\text m,i}>0$)
(similarly, one can model specific consumer ($P_{\text m,i}<0$) behaviour).
On the other hand, the network's topology can be varied by means of the susceptance matrix' coefficients $\{B_{ij}\}_{i,j=1,..,N}$ to search for topological 
aspects, that favour grid stability and investigate the impacts of progressive grid decentralization.

\begin{acknowledgments}

\noindent{\bf ACKNOWLEDGMENTS}\\
\\ 
My co-authors Joachim Peinke, Oliver Kamps, and me owe special thanks to Rudolf Friedrich, who passed away too soon in August 2012. This research work was mainly done in the course of my diploma thesis. Being a member of his working group, I could benefit from his broad stock of knowledge as well as his physical intuition.
We will always keep in good memory his friendly manner and open-mindedness. K. S.
\end{acknowledgments}

\bibliography{Bibliothek/bibliothek}

\end{document}